\documentclass[%
twocolumn,
 amsmath,amssymb,
 aps,
]{revtex4-2}

\usepackage{CJKutf8}% For typing CJK characters on Overleaf
\usepackage{graphicx}% Include figure files
\usepackage{dcolumn}% Align table columns on decimal point
\usepackage{bm}% bold math
\usepackage{import, comment, ulem}
\usepackage{textcomp} % disable warning from gensymb package, must be imported before gensymb
\usepackage{gensymb, commath, empheq, mathtools, stmaryrd, mathrsfs}
\usepackage{graphicx, float, xcolor}%caption, subfig, 
\usepackage[caption=false]{subfig} %revtex is not compatible with subfig captions
\usepackage[toc,page]{appendix}
\usepackage{url, hyperref}
\newcommand{\pvec}[1]{\vec{#1}\mkern2mu\vphantom{#1}}
\newcolumntype{C}[1]{>{\centering\arraybackslash}p{#1}}
\CJKfamily{gbsn}

\begin{document}
%\begin{CJK*}{}{}
\begin{CJK*}{UTF8}{gbsn}
\preprint{APS/123-QED}

\title{WISPLC: Search for Dark Matter with LC Circuit}% Force line breaks with \\
%\thanks{A footnote to the article title}%

\author{Zhongyue Zhang (张钟月)}%0000-0003-2631-5345
\affiliation{%
 Institut f\"ur Experimentalphysik, Universit\"at Hamburg, Luruper Chaussee 149, 22761 Hamburg,  Germany
}%
 %\altaffiliation[Also at ]{Institute of Experimental Physics, Hamburg University.}%Lines break automatically or can be forced with \\
\author{Oindrila Ghosh}%0000-0003-2226-0025
\affiliation{%
 II. Institut f\"ur Theoretische Physik, Universit\"at Hamburg, Luruper Chaussee 149, 22761 Hamburg,  Germany
}%
 %\email{Second.Author@institution.edu}
 \author{Dieter Horns}%0000-0003-1945-0119
\affiliation{%
 Institut f\"ur Experimentalphysik, Universit\"at Hamburg, Luruper Chaussee 149, 22761 Hamburg,  Germany
}%

\date{\today}% It is always \today, today,
             %  but any date may be explicitly specified

\begin{abstract}
The focus on dark matter search has expanded to include low-mass particles such as axions or axion-like particles, and novel theoretical schemes extending the phenomenological landscape, within QCD and beyond, also garnered additional interest in recent decades. Assuming dark matter is composed of axions, in presence of a solenoidal magnetic field, they induce a displacement current that gives rise to a toroidal magnetic field. The Weakly Interacting Slender Particle detection with LC circuit (WISPLC) is a precision direct detection experiment that will search for light dark matter candidates such as axion-like particles in parts of the parameter space previously unexplored. We present two detection schemes of the signal in a pickup loop capturing the flux of this toroidal magnetic field. WISPLC operates in a broadband and a resonant scheme where a LC circuit is used to enhance the signal with an expected Q factor $\sim 10^4$. Taking into account the irreducible flux noise of the detector, we estimate the sensitivity of the experiment in the axion mass range between $10^{-11}$~eV and $10^{-6}$ eV to reach a detectable axion-photon coupling of $g_{a\gamma\gamma}\approx 10^{-15}~\mathrm{GeV}^{-1}$, making it possible to probe mass ranges corresponding to ultralight axions motivated by string theory. The WISPLC experiment is fully funded and currently in the construction phase.
\end{abstract}

%\keywords{Suggested keywords}%Use showkeys class option if keyword
                              %display desired
\maketitle
\end{CJK*}

\section{\label{sec:level1}Introduction}

Axions are light pseudo-scalar Goldstone bosons that were proposed as a solution to the strong CP problem in quantum chromodynamics (QCD) via the $U(1)_\text{PQ}$ Peccei-Quinn (PQ) symmetry breaking \cite{PQ} \cite{weinberg} \cite{wilczek} and also arise in String theoretical frameworks which, in particular, motivate ultralight species \cite{axiverse}. In recent decades, axions have moved to the forefront of exploring physics beyond the Standard Model, primarily as a potential dark matter candidate.  

The mass of axions $m_{a}$ relates to the PQ scale $f_a$ as:
\begin{equation}
    m_a^2 f_a^2 \approx m_\pi^2 f_\pi^2 % \  \frac{m_u m_d}{(m_u + m_d)^2}
    \label{qcdaxion}
\end{equation}
where $m_\pi$ and $f_\pi$ are the mass and decay constant of pions, respectively. Axion-like particles (ALPs) are very similar to QCD axions in terms of their properties but are not constrained by Eq. \ref{qcdaxion}, and therefore, can take a broad range of parameters. Axions interact with photons via the interaction term: 
\begin{equation}
    \mathcal{L} = \frac{1}{2}(\partial_\mu a)^2 - \frac{1}{2}m_a^2a^2 - \frac{1}{4}F_{\mu\nu}\tilde{F}^{\mu\nu} + \frac{1}{4}g_{a\gamma\gamma}aF_{\mu\nu}\tilde{F}^{\mu\nu},
\end{equation}
where $a$ is the axion field and $F_{\mu\nu}$ and $\tilde{F}_{\mu\nu}$ are respectively the electromagnetic field tensor and its dual. The axion-photon coupling strength $g_{a\gamma\gamma}$ is model-dependent,
\begin{equation}
    g_{a\gamma\gamma} = \frac{\alpha}{2\pi f_a} \bigg[\frac{E}{N}-1.92(4)\bigg],
\end{equation}
where $\alpha$ is the fine structure constant, and $E$ and $N$ are respectively the electromagnetic and colour anomaly coefficients dependent on the choice of axion models such as DFSZ-I, DFSZ-II, and KSVZ, for which $E/N$ equals to $8/3$,  $2/3$, and $0$ respectively \cite{dine}\cite{dine1980phys} \cite{Zhitnitsky:1980tq} \cite{kim1979weak} \cite{shifman1980can}. However, the parameter space window can be further expanded as proposed in \cite{di2017redefining} \cite{di2017window}, and more recently in \cite{DiLuzio2021} and \cite{Sokolov2021}, which extends the search for QCD axions in the nano-electronVolt (neV) range.

In presence of axions, Maxwell's equation is modified as,
\begin{equation}
    \vec{\nabla}\times \vec{B}\ -\ \frac{\partial \vec{E}}{\partial t} \ =\ g_{a\gamma\gamma}\:(\ \vec{E} \times \vec{\nabla}a\ -\ \vec{B}\,\frac{\partial a}{\partial t}\ )\ +\ \vec{j}_{el}, \label{equ:modified_maxwell}
\end{equation}
where $\vec{E}$ and $\vec{B}$ stand for the electric and magnetic field respectively, and $\vec{j}_{el}$ is the electromagnetic current. Equation \ref{equ:modified_maxwell} features the time derivative of the axion field, which is related to the average local axion density as $\langle\rho_a\rangle=\frac{1}{2}\dot{a}^2$.

The characteristic lengthscale of an neV-scale axion is $\sim \mathcal{O}(10^5, \text{m})$, which exceeds the physical dimension of the experiment significantly. Thus, it is safe to assume that axion behaves as a coherent oscillating scalar field in the context of a standard laboratory experiment and we can ignore any spatial variation and assume homogeneity:
\begin{equation}
    a(t) = a_0 \cos{(m_a t)} = \frac{\sqrt{2\rho_{\scriptscriptstyle \text{DM}}}}{m_a} \cos{(m_a t)}, 
    \label{equ:axion_field}
\end{equation}
where $a_0$ is the field amplitude, $\rho_{\scriptscriptstyle \text{DM}} \approx 0.3 \  \text{GeV}/\text{cm}^3$ is the local dark matter density with the assumption $\langle \rho_a \rangle=\rho_{\scriptscriptstyle \text{DM}}$. The combination of Eq. \ref{equ:modified_maxwell} and Eq. \ref{equ:axion_field} shows that the external magnetic field induces an axion-sourced current density $\vec{j}_a$ oscillating at the Compton frequency of axions $\nu_a = m_a c^2/h$, where $h$ is Planck's constant and $c$ is the speed of light. The induced axion current, 
\begin{equation}
    \vec{j}_a(t) = - g_{a\gamma\gamma}\, \vec{B}\ \frac{\partial a}{\partial t},
\end{equation}
subsequently generates a perpendicular toroidal magnetic field $\vec{B}_a$ such that $\vec{\nabla}\times\vec{B}_a=\vec{j_a}$. Due to the random dispersion within the dark matter halo following a Maxwell Boltzmann distribution, the axion signal is expected to have a line profile with a bandwidth $\Delta \nu_a$ centered at the Compton frequency $\nu_a=\omega_a/2\pi=m_ac^2/h$ such that $\Delta \nu_{a} = \nu_{a}\sigma_{\upsilon}^2$ where the dark matter velocity dispersion $\sigma_{\upsilon} = \upsilon/c \sim \mathcal{O}(10^{-3})$ in the Milky Way \cite{Nguyen_2019}. The spread of frequencies of axion can be approximated by the inverse of the so-called axion coherence time $\tau_a$, which is limited by its energy spread $\tau_a \sim 1/m_a\sigma_{\upsilon}^2$. For neV axions, we expect $\tau_a \sim 0.66\,\text{s}$. 

The Weakly Interacting Slender Particle detection with LC circuit (WISPLC) experiment proposes to capture $\vec{B}_a$ with a superconducting loop, amplify it with a LC resonant circuit, and measure it with a Superconducting Quantum Interference Device (SQUIDs) magnetometer. The experiment is built upon the concept of detecting the magnetic field generated by the induced axion current in presence of an external magnetic field, as a means of detecting ultralight axions and ALPs in the local galactic halo.
Besides the basic broadband detection scheme that uses the entire detector bandwidth, it is also possible to integrate a tunable LC circuit in the 
readout to enhance the sensitivity in smaller bandwidths. In this work, we present the idea and design of this haloscope experiment with high-field-strength large-scale superconducting magnets.

\section{Experimental setup for WISPLC}

\begin{figure}[!htp]
\includegraphics[width=1.0\columnwidth]{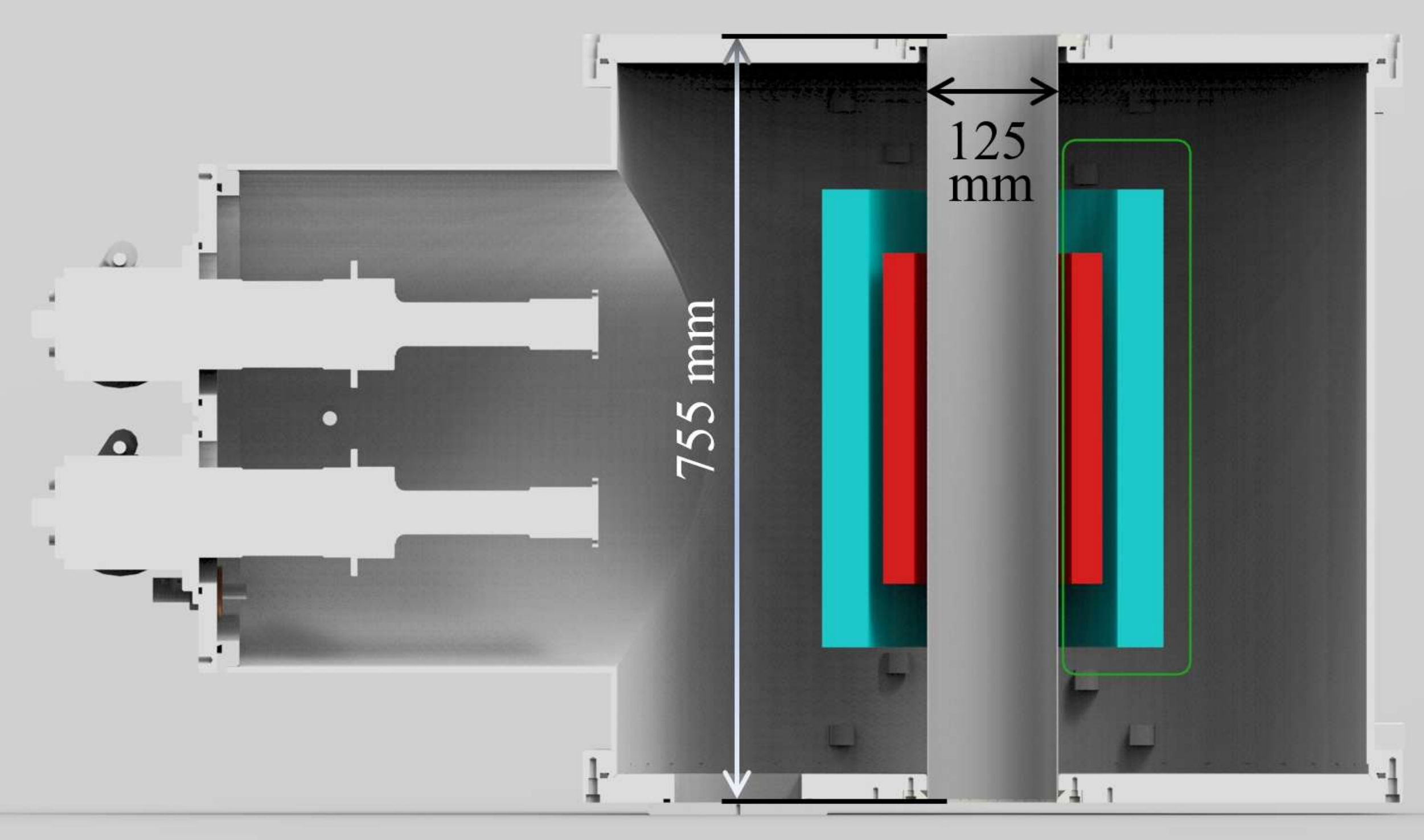}% Here is how to import EPS art
\caption{\label{fig:magnets_step} Model of the large-scale cryogen-free magnet system with a warm bore in the center. The diameter and length of the bore are 125 mm and 755 mm, respectively. Two concentric solenoidal magnets shown in blue and red can produce a maximum magnetic field of 14 T in the center. Shown in green, two individually wired superconducting loops are installed inside the cryostat to enclose the cross-section of the magnets.}
\end{figure}
The key facility of the WISPLC experiment is a large-scale cryogen-free magnet system with warm bore as shown in Fig.~\ref{fig:magnets_step}. The bore has a diameter of 125 mm and a length of 755 mm. Two concentric solenoids, shown in blue and red, wrapped in superconducting wire, can produce a maximum magnetic field of 14 T at the center of the warm bore. Enclosing the cross-section of magnets, two individually wired superconducting loops shown in green is pre-installed inside the cryostat. 

\begin{figure}[!htp]
\includegraphics[width=1.0\columnwidth]{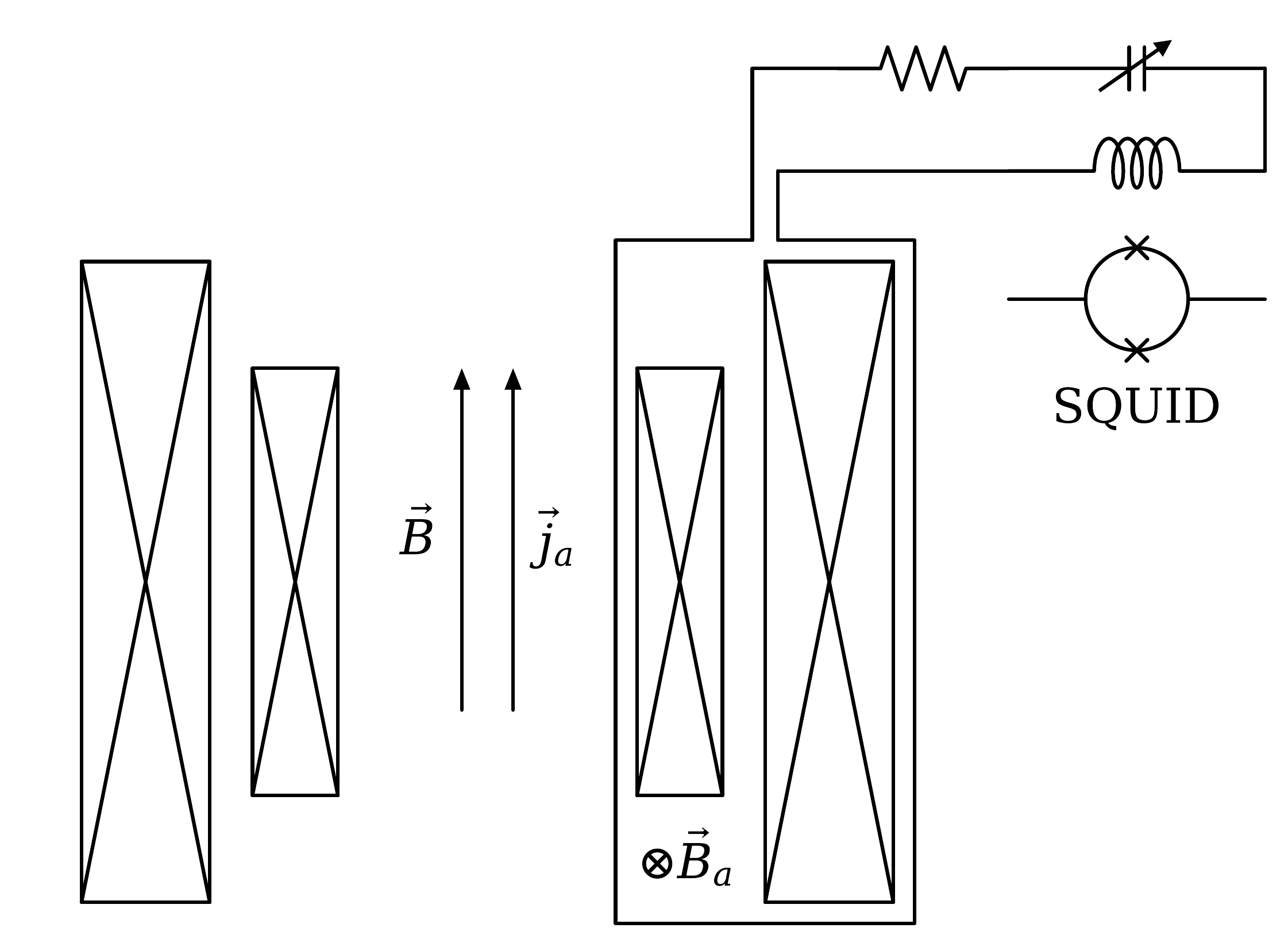}% Here is how to import EPS art
\caption{\label{fig:experiment_setup} Schematics of the proposed experimental setup. The four rectangles with crossings represent the windings of two solenoid magnets. The magnets generate a magnetic field $\vec{B}$ which induces an axion-sourced current density $\vec{j}_a$.}
\end{figure}

A simple experimental scheme is shown in Fig.~\ref{fig:experiment_setup}. The external magnetic field converts axion to an oscillating electric current $\vec{j}_a$, and creates an oscillating toroidal magnetic field $\vec{B}_a$ perpendicular to $\vec{j}_a$. According to Faraday's law, $\vec{B}_a$ induces an AC EMF and subsequently an AC current in the pickup loop, which is then amplified by the tunable LC circuit in the readout, and furthermore converted into magnetic field via an input coil. The SQUID magnetometer will measure the AC current through inductive coupling to the input coil. We can express the magnetic flux passing through the pickup loop as
\begin{equation}
    \Phi_a(t) = g_{\alpha\gamma\gamma} \sqrt{2\rho_{\scriptscriptstyle \text{DM}}}\, C \sin{(m_a t)} \label{equ:axion_flux}
\end{equation}
where $C = | \vec{B}_\text{max}| V_\text{B}$ represents the
 form factor of the experiment and the effective magnetic volume $V_B=\mathcal{G}_\mathrm{V} V_\text{magnet}$ can be interpreted as the equivalent volume with a uniform magnetic field $\vec{B}_\mathrm{max}$ for a given pickup loop geometry. 
The \textit{specific} form factor $\mathcal{G}_\mathrm{V}$ is therefore defined as
\begin{equation}
    \mathcal{G}_V = \frac{1}{|\vec{B}_\mathrm{max}| V_\text{magnet}} \int_\text{loop}\!\! dS \int_{\text{magnet}} \!\!\!\!\! \frac{\vec{B}(\vec{r})\times(\vec{r}-\pvec{r}')}{|\vec{r}-\pvec{r}'|^3} \cdot \hat{n}\,dV   \label{equ:form_factor},
\end{equation} 
identical to the geometric factor in \cite{Kahn2016}. For WISPLC, we estimated the relevant
values through numerical 2D finite element methods to be 
$|\vec{B}_\mathrm{max}|=14$ T, $V_\mathrm{magnet}=0.024~ \mathrm{m}^3$, $\mathcal{G}_\mathrm{V}=0.074$.
We can readily compare with values from other experiments as listed in Table~\ref{tab:form_factor}.
\begin{table}[!ht]
\centering
\caption{Comparison of experimental parameters between WISPLC, ABRA. and SHAFT, $C = |\vec{B}_\text{max}| V_\text{magnet}\,\mathcal{G}_\mathrm{V}$. }
\begin{tabular}{p{0.077\textwidth}>{\centering}p{0.090\textwidth}>
{\centering}p{0.061\textwidth}>
{\centering\arraybackslash}p{0.117\textwidth}>
{\centering\arraybackslash}p{0.099\textwidth}}
\hline\hline
\rule{0mm}{4mm}
 & $|\vec{B}_\text{max}|\,(\text{T})$  & $\,\,\mathcal{G}_\mathrm{V}$ &     $V_\text{magnet}\, (\text{m}^3)$  & $C/C_\text{SHAFT}$ \\[0.1em]
\hline\hline
\rule{0mm}{4mm}
SHAFT\footnotemark[1]      &  1.5           &      0.108\footnotemark[2]       & $9.5\times 10^{-5}$   & $1$\\
\hline
\rule{0mm}{4mm}
\vspace{0.1em}
ABRA.\footnotemark[3]      &  1             &      \hspace{-1.5mm}0.027      & $8.9\times 10^{-4}$   & $1.55$\\
\hline
\rule{0mm}{4mm}
\vspace{0.1em}
WISPLC     &  14            &      \hspace{-1.5mm}0.074      & $2.4\times 10^{-2}$   & $1.60\times10^{3}$\\
\hline
\end{tabular}
\label{tab:form_factor}
\footnotetext[1]{Reference \cite{Gramolin2021}.}
\footnotetext[2]{Estimated from the effective volume $\mathcal{G}_\mathrm{V} = V_B/V$.}
\footnotetext[3]{Reference \cite{Ouellet2019}.}
\end{table}

In Table.\ref{tab:form_factor}, we compare our form factor $C$ with those of ABRACADABRA (ABRA.) \cite{Kahn2016, Ouellet2019, Ouellet2019a, Salemi2021} and SHAFT \cite{Gramolin2021} among other experiments that conduct axion search with SQUID magnetometer with various geometries of magnets and pickup loops. By virtue of the large volume and high magnetic field strength of our magnets, the expected $C$ parameter for WISPLC is about three orders of magnitude larger than those for the other experiments listed.
\begin{figure*}[!htp]
    \subfloat[Readout scheme for broadband detection]{\includegraphics[height=0.20\linewidth]{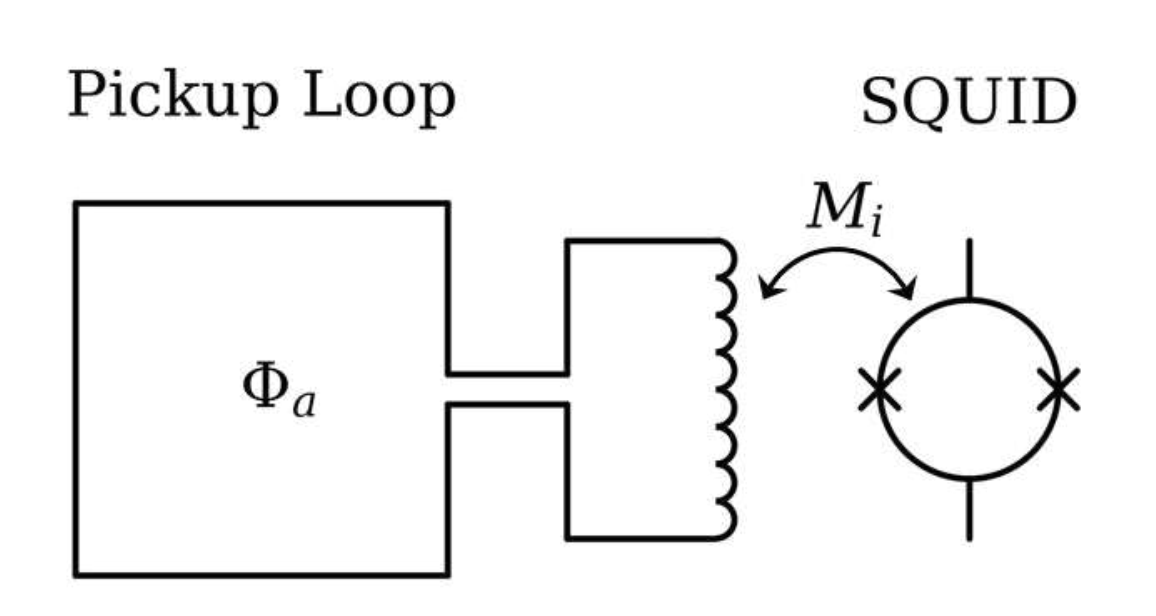}}
    \hspace{10mm}
    \subfloat[Readout scheme with resonant circuit]{\includegraphics[height=0.20\linewidth]{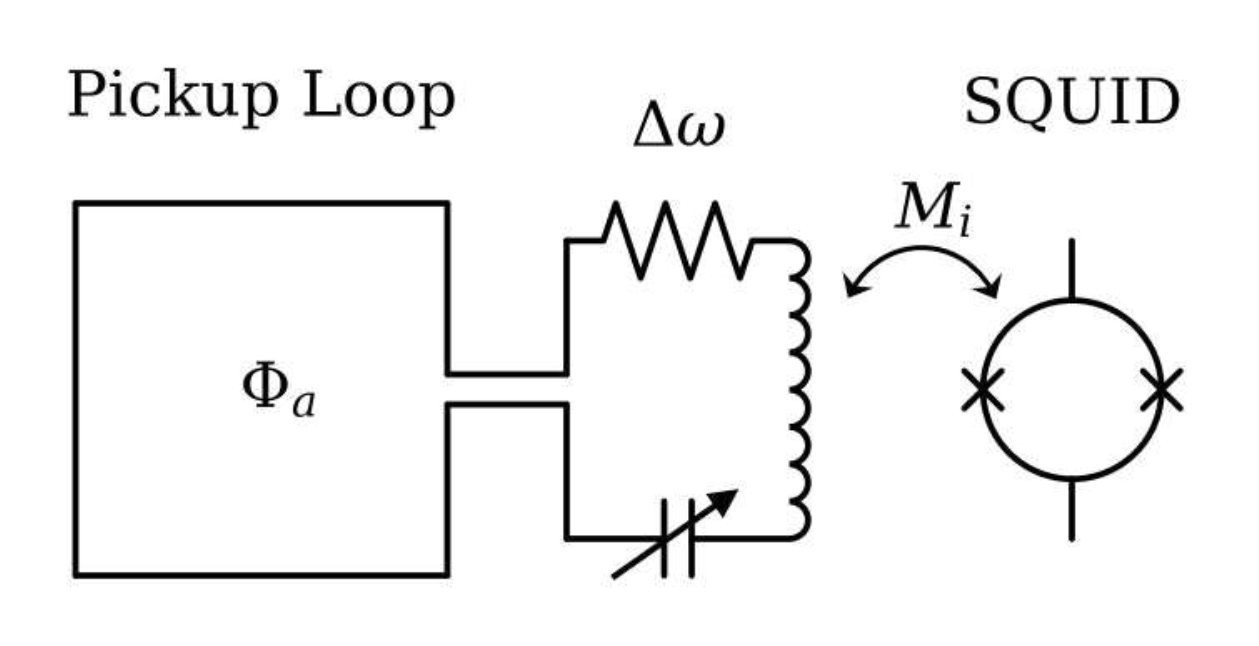}}
    \caption{Two possible readout schemes for axion detection. (a) Readout scheme for broadband detection, the detector is inductively coupled to the pickup loop by an inductor with mutual inductance $M_\mathrm{i}$. (b) Readout scheme with a resonant circuit, the pickup loop is connected to a RLC resonant circuit with bandwidth $\Delta\omega$.}
    \label{fig:readout_scheme}
\end{figure*}
We propose two possible readout schemes with SQUID as shown in Fig.~\ref{fig:readout_scheme}: A. broadband readout, which utilises direct inductive coupling between the pickup loop and SQUID, in which the bandwidth is only limited by the detector and readout electronics, and B. resonant readout, in which a variable LC resonant circuit is inserted between the pickup loop and SQUID and the superconducting current is enhanced by the quality factor $Q$ of the resonant circuit.

\subsection{Broadband Detection}
In the broadband detection scheme, the axion-sourced flux is transferred to the magnetometer through inductive coupling, and the transfer efficiency depends on the properties of the readout circuit, i.e., the mutual inductance $M_\mathrm{i}$ of the coupling and the total inductance $L_\text{sys}$ of the readout circuit:
\begin{equation}
    \Phi_\text{SQUID}(t) = M_\mathrm{i}\, L_\text{sys}^{-1}\, \Phi_a(t)
\end{equation}
$L_\text{sys}$ is the linear addition of the inductances of the pickup loop, coaxial cable, and the coil facing the magnetometer. By virtue of its geometry, the inductance of the pickup loop $\sim 1\, \mu\text{H}$ dominates the total inductance of the system and the flux transfer efficiency $\kappa = \Phi_\text{SQUID}/\Phi_a$ is estimated to be $\sim 4\times10^{-4}.$\\
\indent The bandwidth of this detection scheme is limited by the SQUID readout electronics. A combination of theoretical estimation and laboratory measurement predicts the bandwidth to be around $3.5$ MHz.

\subsection{Resonant Detection}
Tunable LC circuit has the advantage of enhancing signal by a large factor Q at the resonant frequency, however, at the expense of introducing additional noise in the system. Similar strategies have been adopted by ADMX SLIC \cite{Crisosto2020} and ABRA. \cite{Kahn2016, Ouellet2019, Ouellet2019a}). In the resonant detection scheme, the pickup loop is connected to an LC circuit, and the supercurrent oscillating at the frequency $\omega = 1/\sqrt{LC}$ can be enhanced by its quality factor in a restricted bandwidth $\Delta\omega = \omega/Q$. The magnetic flux coupled to SQUID can be approximated as
\begin{equation}
    \Phi_\text{SQUID}(t) \approx Q\, M_\mathrm{i}\, L^{-1}_\text{sys}\, \Phi_a(t)
\end{equation}
In principle, the quality factor as high as $10^{11}$ can be achieved, however, one must consider the problem of oversampling at low frequencies when the bandwidth is significantly smaller than the potential intrinsic signal width. As a trade off, \cite{Devlin2021} picks $Q \sim 4.2(3)\times 10^4$ for the tunable superconducting resonant circuit composed of a toroidal inductor and an electrode, and \cite{Nagahama2016} adopts $Q \sim 10^5$ for a series tunable LC circuit design with four toroidal resonators. For our experiment, we expect a benchmark quality factor of $10^4$. 

\subsection{Setup Summary}
The setup parameters discussed in this section are summarised in Tab~\ref{tab:setup_summary}. 
\begin{table}[!ht]
\centering
\begin{tabular}{C{1.60cm}|C{1.4cm}|C{2.0cm}|C{1.50cm}|C{1.4cm}}
\hline\hline
\rule{0mm}{4mm}
$\vec{B}_\text{max}$ (T)  & $\mathcal{G}_\mathrm{V}$ &      $V_\text{magnets}\,(\text{m}^3)$  & $\kappa$ & $Q$ \\[0.1em]
\hline\hline
\rule{0mm}{4mm}
14                      &      0.074      &    $2.4\times 10^{-2}$ & $4\times 10^{-4}$ & $ 10^4$\\
\hline
\end{tabular}
\caption{Summary of setup parameters of WISPLC experiment. $\vec{B}_\text{max}$ is the maximum magnetic field achievable by the superconducting magnets, $\mathcal{G}_\mathrm{V}$ is the specific form factor characterising the geometric contribution of the setup, $V_\text{magnets}$ is the total volume of the solenoidal magnets; $\kappa$ is the magnetic flux transfer efficiency between the pickup loop and the SQUIDs, and $Q$ is the quality factor of the resonant circuit. }
\label{tab:setup_summary}
\end{table}

\section{Sensitivity estimate}

The typical total flux noise of a SQUID is given by
\begin{equation}
    S_{\Phi} = S_{\Phi,\text{SQUID}} + S_{V, \text{amp}}/V_\Phi^2 + S_{I,\text{amp}}M_\text{dyn}^2
\end{equation}
where $S_{V,\text{amp}}$ and $S_{I,\text{amp}}$ are the current and voltage noise from amplifier, and $V_\Phi$ and $M_\text{dyn}$ are transfer coefficient and current sensitivity of the front end SQUID, respectively. In terms of noise temperature, we can write
\begin{equation}
    S_{\Phi} = 2k_\mathrm{B}(\gamma_VT+T_\mathrm{A})R/V_\Phi^{2}
\end{equation}
where $T$ is the operating temperature of SQUID, $T_\mathrm{A}$ is the effective noise temperature of amplifier (including the parasitic effect of SQUID), $R$ is the shunt resistance of each Josephson junction, and $\gamma_V \approx 8$ \cite{Clarke1979} is the reduced spectral density of the noise voltage $V_\mathrm{N}$ across a ``bare'' SQUID. 
Our SQUID device is an integrated 2-stage current sensor developed by Magnicon \cite{Magnicon} and PTB (Physikalisch-Technische Bundesanstalt) in Berlin, Germany; it consists of an integrated input coil, a single front-end SQUID and a 16 SQUID series array (SSA) as amplifier \cite{Drung2007}. The noise measurement of the shielded sensor at 4 K gives the 1/f corner frequency $\sim$ 4 Hz and white noise floor in terms of flux quanta $\Phi_0$
\begin{equation}
\label{eqn:phi_noise}
    S_{\Phi}^{1/2} \approx 0.9 \ \mu\Phi_0/\sqrt{\text{Hz}}
\end{equation}
which corresponds to an energy resolution of $45$ in units of Planck constant,
and  $\Phi_0=h/(2e)\approx 2\times 10^{-15}\ \text{Wb}$ is the magnetic flux quantum. This preliminary noise floor includes the contribution of vibration noise from a pulse tube cooler; the 14 T superconducting magnets will be cooled with dual GM cooler which is expected to have only slightly larger contribution. It is also expected that the input circuit, untuned or tuned, will contribute additional flux noise in terms of noise temperature $T_\mathrm{N}$. In the following section, we discuss the expected noise performances and expected signal-to-noise ratio (SNR) of the two detection schemes.

\subsection{Broadband sensitivity}

In the broadband detection scheme, a superconducting pickup loop of inductance $L_\mathrm{p}$ connected in series with a pickup coil of inductance $L_\mathrm{i}$ is inductively coupled to a SQUID detector, forming a so-called untuned magnetometer, assumed to be operating at $T$. Using the expression of signal energy for an untuned magnetometer \cite{Clarke1979}, the equivalent noise temperature can be expressed as
\begin{equation}
    T_\mathrm{N} = T \bigg[ 2\gamma_{V\!J} \frac{X_\mathrm{i} V_\Phi M_\mathrm{i}^2\omega}{| Z_\mathrm{T}|^2 R} + \gamma_J \frac{V_\Phi^2 M_\mathrm{i}^4 \omega^2}{| Z_\mathrm{T}|^2 R^2} \bigg] \label{equ:noise_temperature}
\end{equation}
where $X_\mathrm{i} = \omega(L_\mathrm{p} + L_\mathrm{i})$ is the total input reactance, and $Z_\mathrm{T} = \omega^2 M_\mathrm{i}^2/4R_\mathrm{D} + X_\mathrm{i}$ is the total input impedance with contribution from SQUID in the limit $\omega L \ll R_\mathrm{D}$ while assuming the dynamic resistance $2R_\mathrm{D} \approx R$. $\gamma_J\approx 5.5$ is the reduced spectrum density of the noise current $J_\mathrm{N}$ circulating in the SQUID, and $\gamma_{V\!J}\approx 6$ is the correlation coefficient between $V_\mathrm{N}$ and $J_\mathrm{N}$; both values are obtained by digital simulation \cite{Clarke1979}. Combining with parameters of our SQUID detector, we found that $T_\mathrm{N}$ can be treated as independent of frequency and $T_\mathrm{N} \ll T$. Hence, the noise contribution from the ``bare'' SQUID and its amplifiers dominates, and we can safely assume $S_{\Phi, \text{untuned}} \approx S_\Phi$.

\indent The signal-to-noise ratio (SNR) in the frequency domain generally improves with the observation time $t$ as random noises will be cancelled out via averaging; however, the situation becomes confounding when the observation time exceeds $\tau_a$. The solution is to adapt the Dicke radiometer method in order to treat the axion signal as bandwidth-limited. The SNR can then be summarised as follows: 
\begin{equation}
\text{SNR} \;\; = \;\;
    \begin{dcases}
         \;\; \frac{\Phi_\text{SQUID}}{S_{\Phi}^{1/2}}\;t^{1/2}, \;\;\;\;            &  t \leq \tau_a\\
         \;\; \frac{\Phi_\text{SQUID}}{S_{\Phi}^{1/2}}\;(t\,\tau_a)^{1/4} , \;\;\;\;            &  t > \tau_a
    \end{dcases}
    \label{eq:snr}
\end{equation}
For a $2\sigma$ detection of neV-scale axions, this leads to an estimated sensitivity for the axion-photon coupling of

\begin{align}
\begin{split}
g_{a \gamma \gamma, 2\sigma}& \gtrsim {} 8 \times 10^{-13} \, \mathrm{GeV}^{-1}             \left(\frac{m_{a}}{10^{-9}\,\mathrm{eV}}\right)^{1/4} \left(\frac{\sigma_{v}}{10^{-3}}\right)^{1 / 2} \\
& \left(\frac{\rho_{\scriptscriptstyle \text{DM}}}{0.3\, \mathrm{GeV} / \mathrm{cm}^{3}}\right)^{-1 / 2} \left(\frac{\kappa}{4\times10^{-4}}\,\frac{C}{0.025\,\mathrm{m}^3\mathrm{T} }\right)^{-1} \\
& \left(\frac{t}{100 \text { days }}\right)^{-1 / 4}\left(\frac{S_{\Phi}^{1 / 2}}{0.9\, \mu \Phi_{0} / \sqrt{\mathrm{Hz}}}\right)
\end{split}
\end{align}

\begin{figure}[!htp]
\includegraphics[width=1.0\columnwidth]{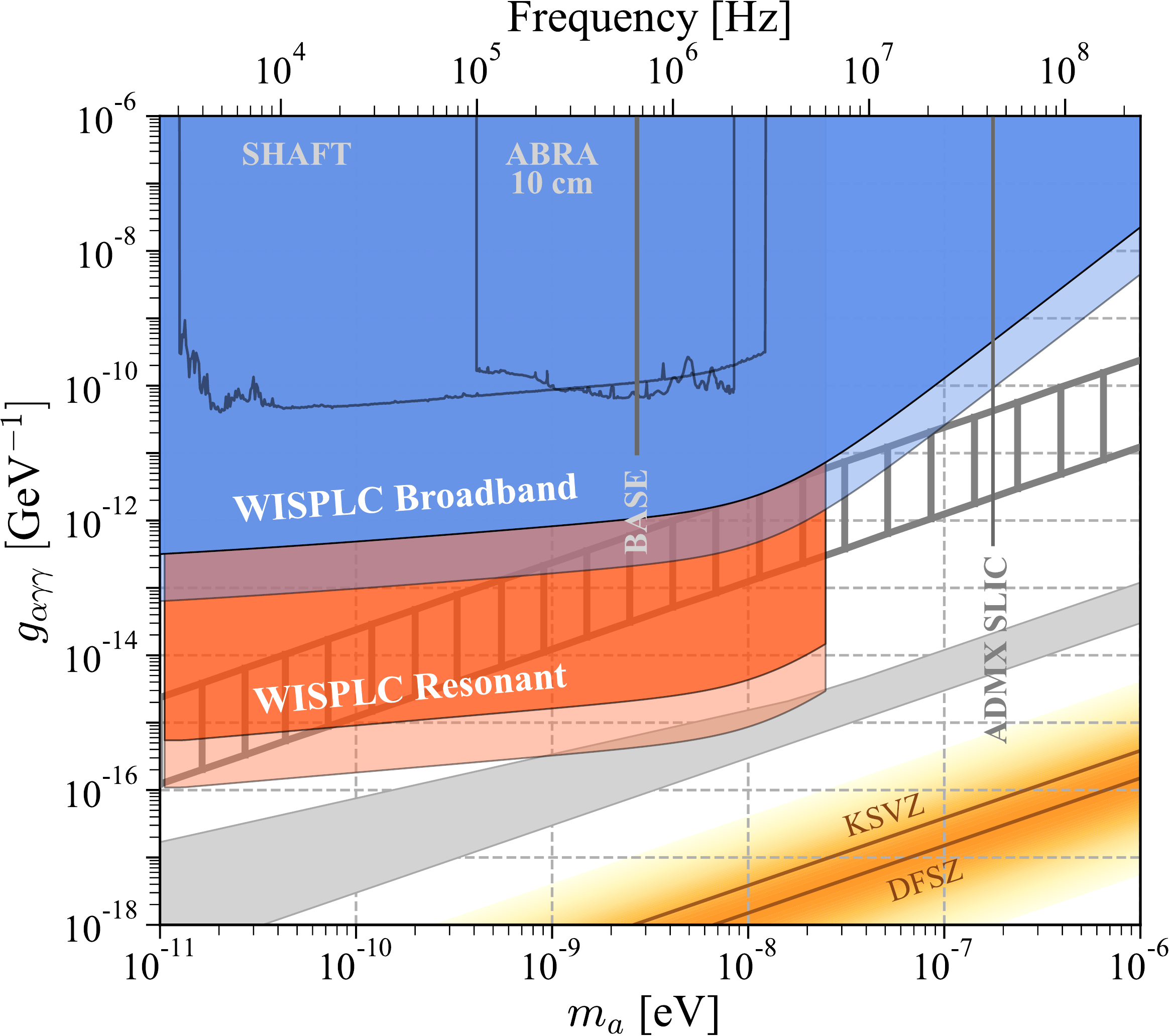}% Here is how to import EPS art
\caption{\label{fig:sensitivity} Anticipated  2$\sigma$ exclusion limit in the parameter space for both broadband (blue) and resonant (orange) detection schemes. The light-shaded blue and orange areas represent the sensitivity with improved read-out. The total measurement time is 100 days for both schemes. In the resonant scheme,
the scan with bandwidth $\omega/Q$ (2 min for each frequency with 1 min of effective integration time) will cover the mass range from $10^{-11}$ eV to $2.5\times 10^{-8}$ eV. The favoured parameter space of the trapped misalignment axion model \cite{DiLuzio2021} and the photophilic axion model \cite{Sokolov2021} are shown in light-shaded grey area and dark grey lines, respectively.}
\end{figure}

In Fig.~\ref{fig:sensitivity}, 
axions with mass and coupling in the blue area would be 
detectable with WISPLC at 2$\sigma$ after an integration time of $t_\mathrm{BB} = 100$ days. The parameter spaces of two newly proposed frameworks that extend the QCD axion band, photophilic axions \cite{Sokolov2021} and trapped misalignment \cite{DiLuzio2021}), are plotted in dark grey lines and light-shaded grey area, respectively. Due to the limitation of bandwidth of the read-out to 
3.5~MHz, the sensitivity deteriorates
quickly for axion masses larger than
$\approx10$~neV. Given that
the sensitivity is not
sufficient to cover the 
expected coupling for some
of the theoretical models for axionic dark matter, 
we consider several ways to 
improve the sensitivity: (i) longer integration time,
(ii) increased form factor, (iii) improved flux transfer efficiency $\kappa$, (iv) decreased noise, and (v) resonantly 
tuned read-out (see next
section). The most promising approach is to 
increase the coupling. This
could be achieved by e.g., 
increasing the geometrical size of the SQUID loop,
ultimately exchanging the pickup loop with a SQUID. In Fig.~\ref{fig:sensitivity}, the sensitivity of an improved read-out with a
scaled coupling of $\kappa=2\times10^{-3}$ is shown in light-shaded blue. This
improved sensitivity is sufficient to constrain the aforementioned dark matter models.

\subsection{Sensitivity of a tuned circuit}
In case of the resonant detection scheme, the search would be
limited to a narrow frequency band around the resonant
frequency. The bandwidth of the search would be $\Delta \nu \propto 1/Q$, where $Q$ is the quality factor of the circuit. Considering a classical design of a tuned magnetometer consisting of input capacitance $C_\mathrm{i}$, input inductance $L_\mathrm{p}+L_\mathrm{i}$ and circuit dissipation $R_\mathrm{i}$, the equivalent noise temperature can be analysed similarly as Eq.~\ref{equ:noise_temperature} with extra thermal noise coming from $R_\mathrm{i}$ and $S_{\Phi, R_\mathrm{i}}(\nu) = 4k_\mathrm{B} T R_\mathrm{i} M_\mathrm{i}^2/| Z_\mathrm{T}|^2$ \cite{Clarke1979}. We corroborate that $S_{\Phi, R_\mathrm{i}}$ significantly exceeds the intrinsic SQUID noise and dominates the input noise contribution. However, with novel LC circuit design consisting of a toroidal superconducting inductor and an electrode in parallel \cite{Devlin2021}, the voltage noise level at resonant frequency of 0.7 MHz is demonstrated to be as low as -90 dB$V_\mathrm{pp}$, which is negligible compared to the intrinsic noise level of the SQUID. Further optimisation might be achieved with impedance matching circuit \cite{chaudhuri2021optimal}. Here we assume optimal noise performance of the resonant circuit $S_{\Phi, \text{tuned}} \approx S_\Phi$.

For the resonant detection, we scan between $10^{-11}$~eV and $2.5\times10^{-8}$ eV, and assume a 1-min interval for the tuning of the LC circuit which leaves $t_\mathrm{Res} \approx 1$ min integration time for each frequency scan. For $m_a \geq 10^{-11}$ eV, the total enhancement on the SNR compared to the broadband detection is
\begin{equation}
    Q_\mathrm{Res} \approx Q\bigg(\frac{t_\mathrm{Res}}{t_\mathrm{BB}}\bigg)^{1/4} \approx 515
\end{equation}
In Figure~\ref{fig:sensitivity} the orange area shows the 2$\sigma$ exclusion limit of all resonant bands, and the light-shaded orange area has an improved flux transfer efficiency similar to the light-shaded blue area.

\section{Summary and outlook}

We have presented an experiment, WISPLC, 
to search for axions as dark matter in the mass range up to 
tens of neV. WISPLC is currently under construction.
The foreseen sensitivity to the two-photon coupling
of axions is improving on existing experimental bounds by two orders
of magnitude. The improvement is primarily due to a larger and stronger
magnet in comparison to previous experiments. 
The sensitivity will be reached with a conventional SQUID broadband read-out. 
A tuned version of the read-out will improve the sensitivity by up to three orders of magnitude in comparison to the broadband readout. This way, we will be sensitive to the detection of axion dark matter with an axion-photon coupling as in recent theoretical proposals \cite{Sokolov2021,DiLuzio2021}. 
The sensitivity of this setup will not be sufficient to reach the canonical QCD axion range of couplings.

\begin{acknowledgments}
We thank Ciaran O'Hare for useful comments. This project is funded by the Deutsche Forschungsgemeinschaft (DFG, German Research Foundation) under Germany's Excellence Strategy – EXC 2121 ``Quantum Universe'' – 390833306 
and through the DFG funds for major instrumentation (s/c magnet) DFG INST 152/824-1.
\end{acknowledgments}

\bibliographystyle{apsrev4-2}
\bibliography{WISPLC}

\end{document}